\documentclass[aps,pra,twocolumn,floatfix]{revtex4-1}
\usepackage{amsmath,amssymb}
\usepackage{graphicx}
\usepackage{graphics}
\usepackage{epsfig}
\usepackage{color}
\usepackage{xcolor}
\usepackage{bm}
\usepackage[latin9]{inputenc}
\usepackage{ulem} 

\begin{document}
\title{Reversible ultrafast soliton switching in dual-core highly nonlinear optical fibers}
\author{Nguyen Viet Hung$^1$, Le Xuan The Tai$^2$, Ignac Bugar$^{3,4}$, Mattia Longobucco$^{3,5}$, Ryszard Buczynski$^{3,5}$, Boris A. Malomed$^{6,7}$ and Marek Trippenbach$^5$}
\affiliation{$^1$Advanced Institute for Science and Technology,
Hanoi University of Science and Technology, Hanoi, Vietnam.
    \\
$^2$Faculty of Physics, Warsaw University of Technology, PL-00662 Warsaw, Poland.
    \\
$^3$Department of Glass, Lukasiewicz Research Network Institute of Electronic Materials Technology, Wolczynska 133, 01-919 Warsaw, Poland.
    \\
$^4$International Laser Centre, Ilkovicova 3, 841-04 Bratislava, Slovakia.
\\
$^5$Faculty of Physics, University of Warsaw, ul. Pasteura 5, PL--02--093 Warszawa, Poland.
\\
$^6$Department of Physical Electronics, School of Electrical Engineering, Faculty of Engineering, and Center for Light-Matter
Interaction, Tel Aviv University, Tel Aviv 69978, Israel.
\\
$^7$Instituto de Alta Investigaci\'{o}n, Universidad de Tarapac\'{a}, Casilla 7D, Arica, Chile.}

\begin{abstract}
We experimentally investigate a nonlinear switching mechanism in a dual-core highly nonlinear optical fiber. We focus the input beam of
femtosecond pulses on one core only, to identify transitions between inter-core oscillations, self-trapping in the cross core, and self-trapping of the pulse in the straight core. A model based in the system of coupled nonlinear Schr\"{o}dinger equations provides
surprisingly good agreement with the experimental findings.
\end{abstract}

\maketitle
\label{sec:intro} 
Realization of all-optical switching in a simple
format has long been a challenge for nonlinear fiber optics. The
concept of nonlinear directional couplers based on dual-core fibers
was introduced theoretically in early
1980s~\cite{SSFM,Jensen1982,Maier1982}. Since then considerable
efforts were devoted to {the characterization} and optimization of
the device performance \cite{RH,GPA}. {In particular}, a promising
demonstration of ultrafast nonlinear switching had been reported
utilizing femtosecond pulses in {the
normal-group-velocity-dispersion (GVD) range} of the silica-fiber
coupler~\cite{Friberg88}. The main limitations of ultrafast
nonlinear switching in conventional nonlinear couplers are
relatively high powers ({\symbol{126}100} kW) required for the
signal redirection, and {the ensuing }breakup in the {temporal
domain}~\cite{Friberg88,Steg}. Additionally, the switching
performance is compromised by the {intra-channel and inter-modal}
GVD, {which strongly affects} pulses of width \symbol{126} $100$ fs.
To avoid the degradation driven by these factors, it was proposed to
exploit temporal solitons~\cite{Trillo}, taking advantage of their
robustness. Numerous theoretical
works~\cite{Romagnoli,Kivshar,Chu1995,Chiang,Uzunov,Driben} reported
diverse schemes of the soliton switching.

Despite {the }theoretical advances, {very few experimental studies have been
performed for switching of temporal solitons in nonlinear couplers}, with
results remaining far behind the theoretical predictions. The experimental
works, exploiting the soliton propagation in dual-core photonic-crystal
fibers ({PCFs}) ~\cite{Betlej2006,Stajanca2014}, {were hampered by the
fission} of naturally emerging higher-order solitons, resulting in output
distributed chaotically between the two channels \cite{Herrmann2002}. Later,
an extensive numerical study for an air-glass dual-core PCF made of a highly
nonlinear lead silicate glass (PBG-08), {had revealed} a possibility of
self-trapping of higher-order solitons, {following} their self-compression
\cite{Stajanca2016}. Such an effect, which tends to keep {a }spectrally
broadened pulse in one fiber core, was {demonstrated }in a multichannel
fiber structure, {as a basis of the creation of \textquotedblleft arrayed }%
light bullets" \cite{Minardi2010}. Motivated by these concepts, a new study {%
of self-trapping, alternating }between the two fiber cores, {was initiated,
aiming at achieving high-contrast} switching performance. {It is focused on}
{the performance of} a highly-nonlinear dual-core fiber with {two soft glass
kernels}. Strong nonlinearity is ensured by using the PBG-08 glass, while
the complex air-glass PCF structure is replaced by a low-index glass \cite%
{Longobucco2019}. The high-index contrast ($0.4$) between the core and cladding in
this system supports very {efficient} switching performance, as predicted by
simulations \cite{Longobucco2020a}. Moreover, a higher level of the
dual-core symmetry was achieved in this fiber, in comparison {to previously
used dual-core PCFs, which is necessary for the operation of all-optical
switching in dual-core fibers }\cite{Curilla2018}.

The pilot experiments in the optimized dual-core-fiber, using{\ pulses of} $%
100${-fs duration, with carrier wavelength} $1700$ nm, {have} demonstrated, {%
for first time, high-contrast} (16.7 dB) {switching} in {the }soliton regime
\cite{Longobucco2020b}. In the present paper {we report essentially more
advanced} experimental results achieved in the C-band (at $1560$ nm), {using
a} new generation of {strongly nonlinear high-index-contrast dual-core fibers%
}. The experimental findings are {supported by} simulations {which use a
model with experimentally relevant parameters}.

\label{sec:TM}The model is based on the system of linearly-coupled nonlinear
Schr\"{o}dinger equations (NLSEs){\ }\cite{Liu2010,Zhao2010,Li2014,review},
written for complex envelopes $\Psi _{1,2}\left( t,z\right) $ of
electromagnetic waves in the cores
\begin{equation}
\begin{array}{c}
i\partial _{z}\Psi _{1}=-(1/2)\beta _{2}\partial _{t}^{2}\Psi _{1}-\gamma
|\Psi _{1}|^{2}\Psi _{1}-\kappa \Psi _{2}, \\[1mm]
i\partial _{z}\Psi _{2}=-(1/2)\beta _{2}\partial _{t}^{2}\Psi _{2}-\gamma
|\Psi _{2}|^{2}\Psi _{2}-\kappa \Psi _{1},%
\end{array}
\label{eq:fundament}
\end{equation}%
where $z$ and $t$ are the propagation distance and time in physical units,
with $\beta _{2}$, $\gamma $, and $\kappa $ representing, respectively, the
GVD, Kerr nonlinearity, and inter-core coupling. By means of rescaling, $t=%
\sqrt{\beta _{2}/\kappa }\tau \equiv t_{0}\tau $, $z=\zeta /\kappa $, $\Psi
_{1,2}=\sqrt{\kappa /\gamma }A_{1,2}$, Eqs. (\ref{eq:fundament}) are cast in
the normalized form, with $\beta _{2},\gamma $, and $\kappa $ set equal to $1
$. The model does not include effects of secondary importance, such as the
Raman scattering, self-steepening, and higher-order GVD effects (cf. Ref. \cite%
{Longobucco2020b}), as simulations of the extended system demonstrate no
essential differences, in terms of switching and retention in the straight channel, except for a
uniform factor adjusting the experimental and theoretical energy scales,
which is $E_{\text{exp}}\approx 1.25E_{\text{theor}}$, as produced by Eqs. (%
\ref{eq:fundament}), see below.

Simulations of this system were run with the input corresponding to the
experiment performed in this work, \textit{viz}., a soliton-like pulse with
independent amplitude $a$ and inverse width $\eta $ \cite{Satsuma1974},
coupled at $\zeta =0$ into one (\textit{straight}) channel:%
\begin{equation}
A_{1}(0,\tau )=a\,\mathrm{sech}(\eta \tau ),\,\,A_{2}\left( 0,\tau \right)
=0,  \label{eq:initpulse}
\end{equation}%
the FWHM width of the pulse being $t_{\mathrm{FWHM}}=1.76t_{0}/\eta $.

\begin{figure}[th]
\begin{center}
\includegraphics[width=\linewidth]{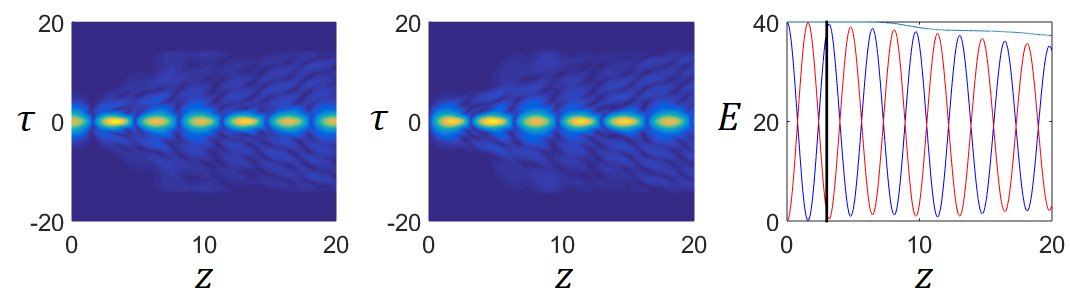} \includegraphics[width=%
\linewidth]{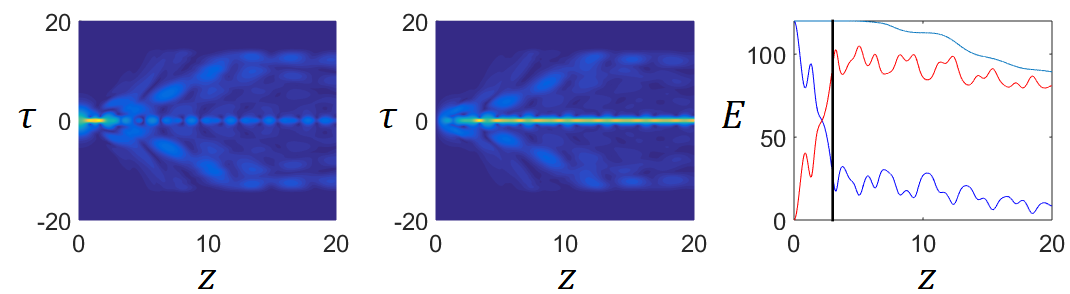} \includegraphics[width=\linewidth]{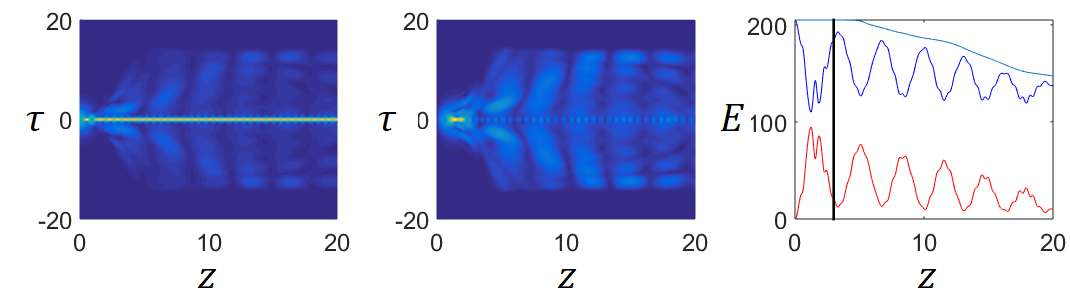}
\end{center}
\caption{The switching between channels, produced by simulations of Eq. (%
\protect\ref{eq:fundament}) for three values of the input amplitude (\protect
\ref{eq:initpulse}), $a=1.15$, $2.0$, $2.6$ (from top to bottom), and a
fixed inverse width, $\protect\eta =0.78$. The left and central columns
display spatiotemporal patterns of the intensities, $\left\vert A_{1,2}(z,%
\protect\tau )\right\vert ^{2}$, in the straight and cross channels,
respectively. The blue and red curves in the right column show the energy in
each channel (and the total energy, shown by the cyan curve) vs. the
propagation distance. The top, central, and bottom panels represent,
severally, regimes with periodic inter-core oscillations, self-trapping in
the cross channel, and retention in the straight one, respectively. The
vertical line at $\protect\zeta =3$ denotes the length of the fiber in the
experiment.}
\label{fig:dynamics}
\end{figure}

\begin{figure}[th]
\begin{center}
\includegraphics[width=\linewidth,height=9cm]{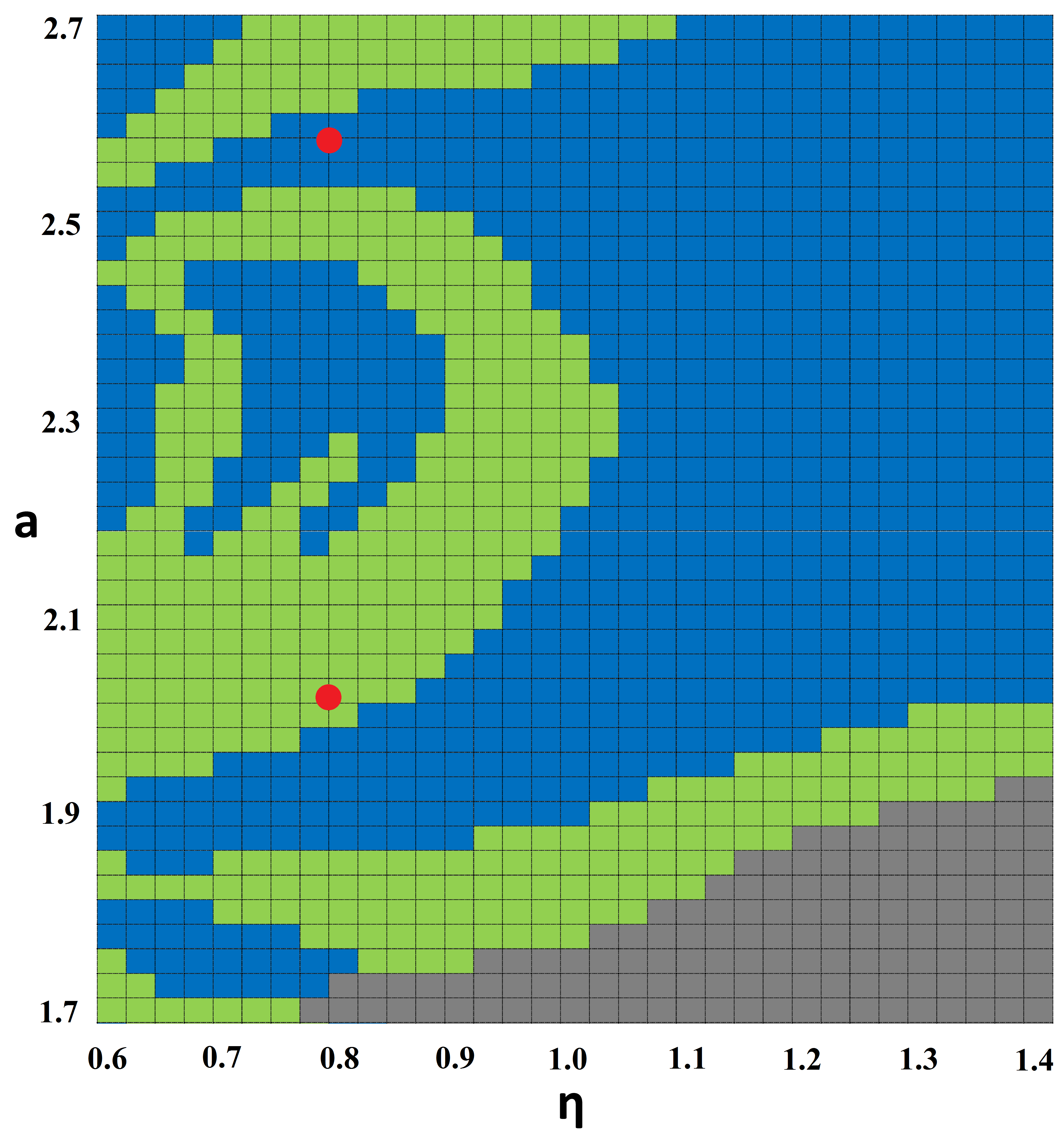}
\end{center}
\caption{The chart of dynamical regimes of the system in the $\left( \protect%
\eta ,a\right) $ plane, produced by simulations of Eqs. (\protect\ref%
{eq:fundament}) with input (\protect\ref{eq:initpulse}). Inter-core
oscillations, switching into the cross channel, and trapping in the straight
one occur in gray, green, and blue areas, respectively. Red circles refer to
two bottom rows in Fig.1. The case shown in the top row falls in the gray
area beneath the frame of the chart.}
\label{fig:table1}
\end{figure}

Simulations of Eqs. (\ref{eq:fundament}) were performed with parameters
corresponding to the all-solid $4.3$ cm long DCF used in the experiment. The
respective parameters, produced by the Lumerical mode solver at wavelength $%
\lambda =1.56\,$ $\mu$m, are the inverse group velocity $\beta
_{1}=6.5610^{-9}$ s/m, $\beta _{2}=-7.73\,\cdot 10^{-26}\,\text{s}^{2}/\text{%
m}$, $\kappa =75\,\text{m}^{-1}$, and $\gamma =0.4\,\,\text{W}^{-1}\text{m}%
^{-1}$ (measured in \cite{Cimek}). Relevant units of the propagation length
and time are $z_{0}=1/\kappa \approx 13$ mm and $t_{0}\approx 32$ fs, with $%
t_{\mathrm{FWHM}}\approx \allowbreak 75$ fs and $\eta =0.78$. In the scaled
notation, the natural period of the population oscillations between the
cores is $\pi $, and the length of the sample is $\approx 3$, making it
appropriate for the realization of the switching. The scaled amplitude in
Eq. (\ref{eq:initpulse}), $a$, is related to the pulse's energy in physical
units
\begin{equation}
E=\left( \kappa /\gamma \right) \tau _{0}\int_{-\infty }^{+\infty
}|A_{1}(0,\tau )|^{2}t_{0}d\tau \approx 1.14a^{2}\kappa t_{_{\mathrm{FWHM}%
}}/\gamma ,  \label{E}
\end{equation}%
For our experimental conditions, this implies $E\approx 30a^{2}$ in units of
pJ.

Input (\ref{eq:initpulse}) in the single channel ($\kappa =0$) generates
intrinsic oscillations of deformed solitons (breathers). According to the
exact solution of NLSE \cite{Satsuma1974}, the (spatial) frequency of the
oscillations is $\omega =4(a-\eta )\eta $ in interval $3/2<a/\eta <5/2$, in
which the breather is a superposition of two fundamental solitons. As a
function of $\eta $, it attains a maximum, $\omega _{\max }=a$, at $\eta =a/2
$.

Including coupling $\kappa $, in Fig.~\ref{fig:dynamics} one observes
interplay between inter- and intra-channel oscillations and emission of
small-amplitude waves in each channel, mostly at the initial stage of the
propagation. For a smaller amplitude of the input, $a=1.15$, i.e.,
relatively weak nonlinearity, we observe quasiharmonic oscillations of the
energy between the cores. At a larger amplitude, $a=2.0$, the nonlinearity
switches the quasi-soliton into the cross channel, where it gets trapped. At
the largest amplitude, $a=2.6$, strong nonlinearity keeps the energy in the
straight channel, with residual oscillations of the radiation between the
channels. Slow decay of the total energy, shown by cyan lines in right-hand
panels, is induced by losses at absorbers, installed at edges of the
time-integration domain, to emulate the radiation loss in the experiment. The losses are negligible for $\protect\zeta \leq3$, which corresponds to the fiber
length in the experiment, $4.3$ cm.

Results of simulations are summarized in Fig.~\ref{fig:table1}, in a chart
of three outcomes in plane $\left( \eta ,a\right) $ of the parameters of
input (\ref{eq:initpulse}), \textit{viz}., periodic oscillations (the gray
area); self-trapping in the cross channel (green), and retention in the
straight one (blue). The increase of $a$ exhibits a natural trend for the
transition of the oscillations into self-trapping in the cross channel,
followed by the transition to the retention of strongly nonlinear pulses in
the straight one. The reverse transition of the self-trapping from the
straight channel to the cross one, with the further increase of $a$,
observed in Fig. \ref{fig:dynamics} at $\eta <1$, is explained by the
interplay of initial solitonic breathing and inter-core oscillations.

\label{sec:ER} The experiments were performed using the above-mentioned dual-core fiber, manufactured and optimized by us which is nearly identical to the one in which the high-contrast soliton switching was demonstrated at the wavelength of 1700nm \cite{Longobucco2020b}. The new fiber differs from the prototype by a larger number (six) of cladding rods in the preform, and by slower drawing speed in a different tower.  The scanning electron microscope (SEM) image of the cross section structure  is  similar  as  was presented in \cite{Longobucco2020b}, only the distance between centers of the two high-index PBG-08 glass cores increased slightly from 3.1 to 3.3 $\mu$m.The recent numerical simulation study took into consideration the real fiber parameters acquired by analysing the SEM image of the DCF cross section. Even though it is noticeable that the best experimental results were obtained at the same fiber length of $4.3$ cm as predicted by the earlier numerical simulation at similar wavelength of $1500$ nm, performed on totally symmetrical hexagonal core structure \cite{Longobucco2019}.

The experiment was run in a setup similar to one in Ref.~\cite%
{Longobucco2020b}. The laser source was a Menlo C-fiber amplified
oscillator, generating $3$ nJ pulses at $1560$ nm, with the pulse width of $%
75$ fs, 
at the repetition rate of $100$ MHz. The output channels of the dual-core
fiber were monitored sequentially by a CCD camera (Electro Optic CamIR1550)
and by an optical spectral analyzer (Yokogawa), collecting spectra
separately from both. Images and spectra were recorded at each level of the
input-pulse's energy.
\begin{figure}[th]
\begin{center}
\includegraphics[width=\linewidth]{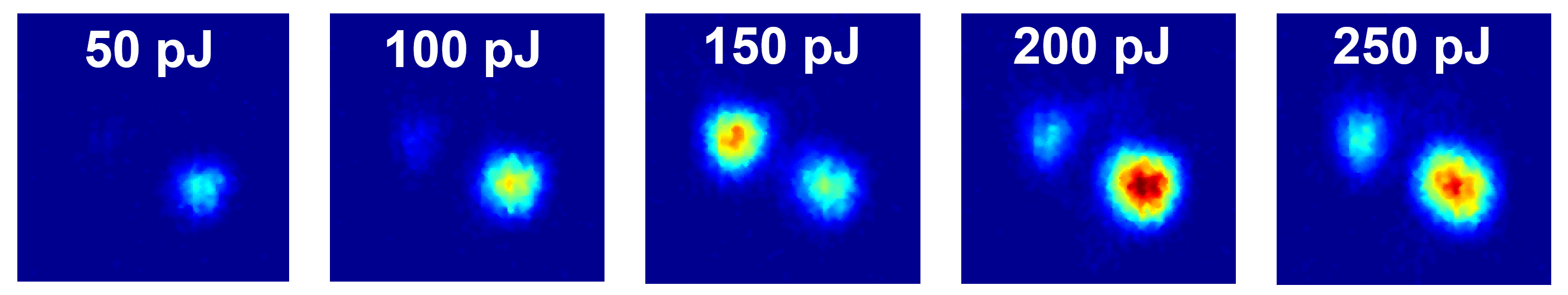}
\end{center}
\caption{A sequence of camera images of the output fiber facet for different
energies of the input.}
\label{fig:table}
\end{figure}
\begin{figure}[th]
\begin{center}
\includegraphics[width=\linewidth]{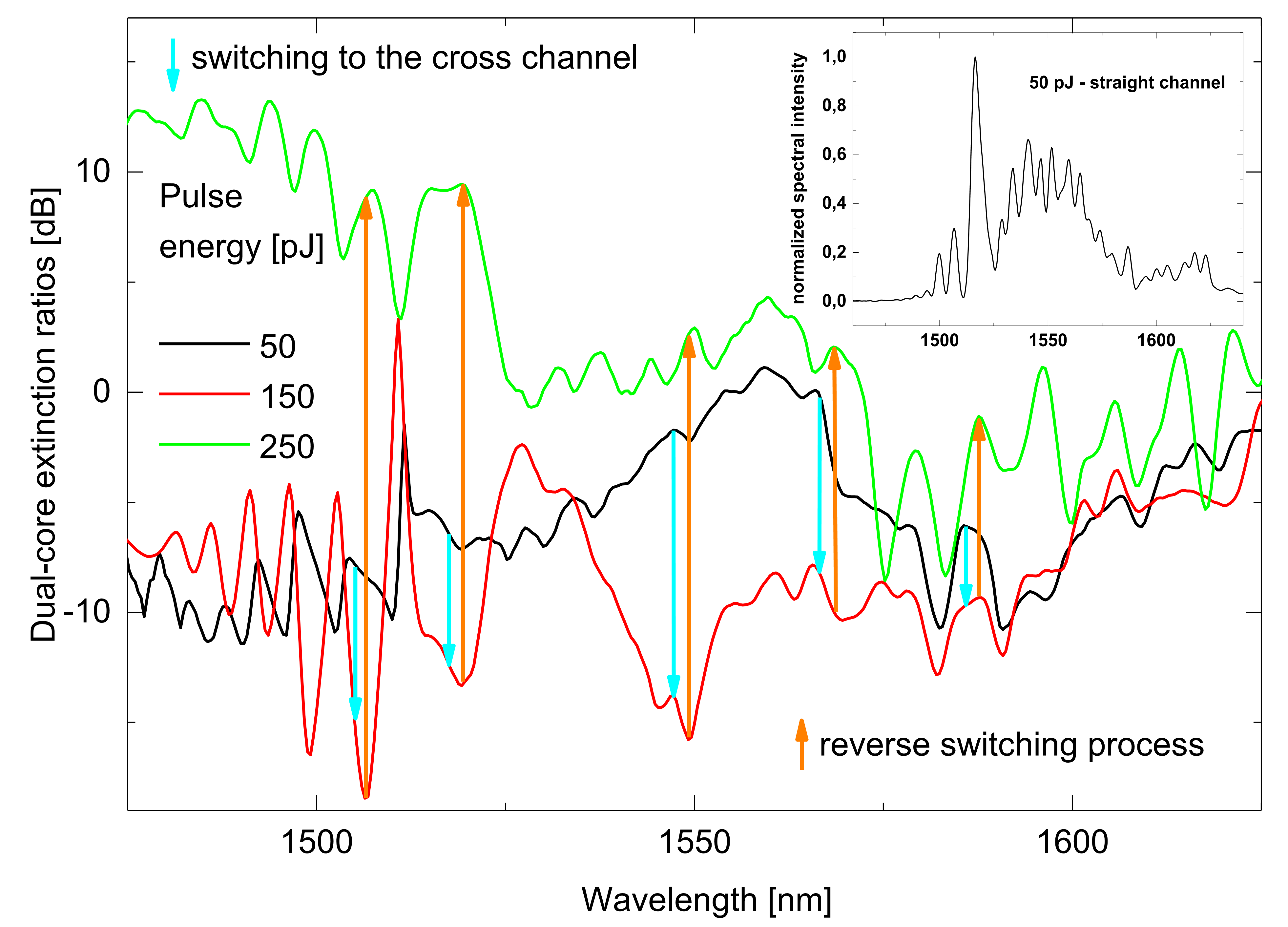}
\end{center}
\caption{The spectrally resolved extinction ratio between output intensities
in the two channels, measured for different input energies, $E$. The inset
displays the respective spectral intensity for $E=50$ pJ. }
\label{fig:table2}
\end{figure}
Figure \ref{fig:table} presents a sequence of camera images recorded with
increasing input energy. These results are similar to the switching
performance reported for the $1700$ nm carrier wavelength \cite%
{Longobucco2020b}, but achieved at lower pulse's energies. The switching
performance includes forth and back switching steps, following the increase
of the energy, at levels $100-150$ pJ and $150-200$ pJ, respectively. They
correspond to the above-mentioned numerically predicted transitions from
oscillations to the trapping in the cross channel, finally followed by the
retention in the straight one.

The spectrally-resolved dual-core extinction ratio, $\mathrm{ER}(\lambda )$,
was calculated, on the basis of the experimental data, using power spectra $%
S_{\mathrm{right}}(\lambda )$ and $S_{\mathrm{left}}(\lambda )$, that were
separately collected from both cores. The dependence of $\mathrm{ER}(\lambda
)\equiv 10\lg \left( S_{\mathrm{right}}(\lambda )/S_{\mathrm{left}}(\lambda
)\right) $ on the input-pulse's energy, $E$, is showed in Fig. \ref%
{fig:table2}, revealing spectral details of the complex switching behavior,
in correspondence with the camera images: at first, $\mathrm{ER}(\lambda )$
decreases with the increase of $E$ in the range of $50-150$ pJ, then it
increases between $150$ and $250$ pJ. The scenario of the all-optical
soliton switching is supported by the fact that only moderate spectral
broadening takes place and the switching is spectrally homogeneous. The same
forth-and-back switching scenario, presented by the arrow pairs in Fig.\ref{fig:table2}, spanning in spectral range of $1510-1575$ nm. This range covers the majority of the pulse energy taking into consideration the basic spectral profile presented in inset of Fig.\ref{fig:table2}.  The forth switching step has lower contrast according to the spectral results comparing the length of the cyan and orange arrows. The origin of this discrepancy is the chromatic aberration of the out-coupling optics avoiding the sharp separation of the two output spectra originating from the straight and cross channel \cite{Longobucco2020b}. Even though, the switching performance reveals clearly a possibility to
direct the energy to either channel in a reversible way. An essential asset
of the operation scheme produced in this work, in the theoretical and
experimental form, is that it provides high switching contrasts without the
requirement of precise adjustment of the fiber length.

As shown in Fig. \ref{fig:table1}, the simulations predict various
possibilities to redirect the soliton between the two channels at $\eta =0.78
$, which corresponds to the pulse width in the experiment. Indeed, varying
amplitude $a$ at fixed $\eta =0.78$, we cross several borders between
regions corresponding to the self-trapping in the cross and straight
channels. This complex structure exists due to the fact that, in the course
of the self-compression, the initial pulse keeps oscillating between the
channels, while the self-trapping occurs only if the soliton acquires a
sufficiently high peak intensity. The pulsations persist in the course of
several periods due to the interplay between the single-channel breathing of
the deformed soliton and inter-core oscillations.

To summarize the comparison between the numerical and experimental results
presented above, we note that the three values of the pulse's amplitude in
Eq. (\ref{eq:initpulse}), $a=1.15$, $2$, and $2.6$, which give rise to the
different outcomes of the transmission through the dual-core fiber,
presented in Figs. \ref{fig:dynamics} and \ref{fig:table1} (quasi-linear
oscillations, self-trapping in the cross channel, and retention in the
straight one), correspond, in physical units, to incident-pulse energies $40$%
, $120$, and $205$ pJ, respectively. The numerical results, obtained for
this set of values of the energy, precisely correspond to the experimental
results observed for energies $50$, $150$, and $250$ pJ, which differ from
their theoretical counterparts by a constant factor, $\approx 1.25$.
Additional effects, such as third-order GVD and Raman effects \cite{SSFM},
account for the remaining discrepancy.

Losses may also affect the soliton propagation in the dual-core fiber.
However, limited fiber lengths, for which the current experiments were
performed, make dissipative effects a relatively weak perturbation,
therefore they are not included in the theoretical model presented above.

In conclusion, reversible high-contrast switching performance of ultrafast
quasi-solitons in the C-band is demonstrated in the strongly nonlinear
dual-core fiber made of soft glass. Both experimental and numerical results
reveal three different scenarios of the soliton propagation, \textit{viz}.,
periodic oscillations, self-trapping in the cross channel, and self-trapping
in the straight one, depending on the energy of the incident pulse. The
experimentally observed scenarios and transitions between them are predicted
by systematic simulations of the system of coupled NLSEs. The results may be
summarized as a well-defined forth-and-reverse soliton-switching effect,
controlled by the monotonous increase of the pulse's energy. Such a
sub-nanojoule high-contrast switching protocol may find applications to the
design of all-optical signal-processing setups. \newline

\textbf{Funding} This work is supported by the Polish National Science
Center through project 2016/22/M/ST2/00261, IB and ML declare support from
project No. 2016/23/P/ST7/02233 under POLONEZ program, which has received
funding from the European Union's Horizon 2020 research and innovation
program under the Marie Sklodowska-Curie grant agreement No 665778. B.A.M.
acknowledges partial support from the Israel Science Foundation through
grant No. 1286/17. N.V.H. was supported by Vietnam National Foundation for
Science and Technology Development (NAFOSTED) under Grant Number
103.01-2017.55. \newline

\textbf{Disclosures} The authors declare no conflicts of interest.

\end{document}